\documentclass[multphys,vecphys]{svmult}

\usepackage{makeidx}     
\usepackage{graphicx}    
\usepackage{multicol}    

\makeindex             

\def 	  \be 		{\begin{equation}}
\def	  \ee		{\end{equation}}

\begin{document}

\title*{Grain Alignment in Molecular Clouds}
\author{A. Lazarian}
\institute{University of Wisconsin-Madison
\texttt{lazarian@astro.wisc.edu}}
%
%
\maketitle
{\bf Abstract} One of the most informative techniques of studying magnetic 
fields in molecular clouds is based on the use of starlight polarization and
polarized emission arising from aligned dust. How reliable
the interpretation of the polarization maps in terms of magnetic
fields is the issue that the grain alignment theory addresses. 
 I briefly review basic physical
processes involved in grain alignment.

\section{Why do we care?}
\label{sec:1}
The fact that the grain got aligned has been known for more than half a 
century. Nevertheless,
it has been always a puzzle why grains get aligned in interstellar medium. 
Very soon after the
discovery of grain alignment by Hall (1949) and Hiltner (1949) it became clear 
that the alignment 
happens in respect to magnetic field. Since that time grain alignment 
stopped to be the issue
of pure scientific curiosity, but became an important missing link of 
connecting polarimetry
observations with the all-important interstellar magnetic fields\footnote{
Additional interest to grain alignment arises from recent attempts to
separate the polarized CMB radiation from the polarized foregrounds (see
Lazarian \& Prunet 2002 for a review).}.

The history of grain alignment ideas is excited (see review by Lazarian 2003) 
and here we do not
have space here to dwell upon it. Within this very short review we will discuss 
the modern understanding
of grain alignment processes applicable to molecular clouds.  Last 
decade has been marked by a substantial progress in understanding new
 physics associated with grain alignment. The theory has become predictive,
which enables researchers to interpret observational data with more confidence.

Recent theory reviews of the grain alignment theory include Roberge (1996), 
Lazarian (2000, 2003). 
Progress in testing theory is covered in Hildebrand (2000), while 
particular aspects of grain dynamics
are discussed in Lazarian \& Yan (2003). The presentation in 
Lazarian (2003) goes beyond molecular cloud
environment and deals with the possibility of alignment in circumstellar 
regions, interplanetary medium,
coma of comets etc. The interested reader may use these reviews to guide 
her in the vast and exciting
original literature on grain alignment.

\section{How do grains rotate?}
Dynamics of grains in molecular clouds is pretty involved
(see Fig.~1). First of all, 
grains rotate. The rotation can
arise from chaotic gaseous bombardment of grain surface and be Brownian, 
or it can arise from 
systematic torques discovered by Purcell (1975, 1979). The most efficient 
among those are torques
arising from H$_2$ formation over grain surface. One can visualize those 
torques imagining a grain
with tiny rocket nozzles ejecting nascent high velocity hydrogen molecules
(see Fig.~1). Grains are known to 
enable atoms of hydrogen to form molecules. The reactions are believed to 
take place over particular
catalytic sites on grain surface. Those catalytic sites act as 
"Purcell rockets". Even when the surroundings
of  dust grains is mostly molecular, grains can rotate suprathermally, i.e. 
with kinetic energies 
much larger that $kT_{gas}$, due to the variation of the accommodation 
coefficient. Indeed, if 
the temperatures of gas and dust are different, those variations allow 
parts of the grain to bounce
back impinging gaseous atoms with different efficiencies. It is easy to 
understand that this also
results in systematic torques. In addition, Purcell (1979) identified 
electron ejection as yet another
process that can drive grain to very large angular velocities.

\begin{figure}
\centering \leavevmode
\includegraphics[height=6cm]{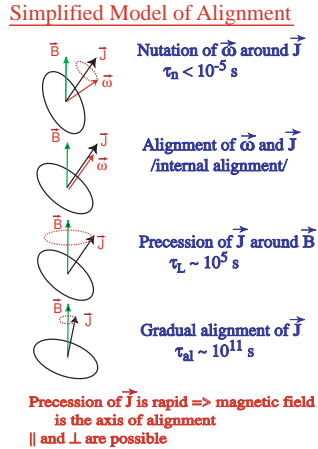}
\caption{Grain alignment implies several alignment
processes acting simultaneously and spanning many time scales
(shown for $10^{-5}$~cm grain in cold interstellar gas). The
rotational dynamics of a grain is rather complex. The internal alignment
introduced by Purcell (1979) was thought to be slower than precession
until Lazarian \& Draine (1999b, henceforth LD99b) showed that it 
happens $10^6$ times
faster when relaxation through induced by nuclear spins is accounted for
(approximately $10^4$~s for the $10^{-5}$~cm grains)}. 
\end{figure}

A very different process of grain spin-up can be found in a very important, 
but not timely appreciated
work by Dolginov \& Mytrophanov (1976).  These authors considered 
differential scattering of
photons of right and left circular polarization by an irregular dust grain. 
As the size of the irregularities
gets comparable with the wavelength, it is natural that interaction of a 
grain with photons will depend
on the photon polarization. Unpolarized light can be presented as a 
superposition of equal number
of left and right circularly polarized photons. Therefore it is clear 
that the interaction with photons
of a particular polarization would deposit angular momentum to the grain. 
The authors concluded that for typical diffuse ISM conditions this process 
should induce grain rotation at 
suprathermal velocities. However, while Purcell's torques became a textbook 
stuff, radiative 
torques had to wait 20 years before they were 
reintroduced to the field (Draine 1996, Draine \& Weingartner 1996, 1997).  

The minimal rotational velocity of grain is the velocity of their Brownian 
motion. This can be 
characterized by a temperature that is somewhere between that of the gas and 
the dust. In the case
of PAHs or very small grains emitting copious microwave radiation as they 
rotate (Draine \& Lazarian 1998)
the effective rotational temperature may be subthermal (see discussion in 
Lazarian \& Yan 2003).

It was realized by Martin (1971) that rotating charged grains will develop
magnetic moment and the interaction of this moment with the interstellar
magnetic field will result in grain precession. The characteristic
time for the precession was found to be comparable with $t_{gas}$. 
However, soon  a process that
renders much larger magnetic moment was discovered (Dolginov \& Mytrophanov 
1976). This process is the 
Barnett effect, which is converse of the Einstein-de Haas effect.
If in Einstein-de Haas effect a paramagnetic body starts rotating
 during remagnetizations
as its flipping 
electrons transfer the angular momentum (associated with their spins)
 to the
lattice, in the Barnett effect
the rotating body shares its angular momentum with the electron
subsystem  causing magnetization. The magnetization
is directed along the grain angular velocity and the value
of the Barnett-induced magnetic moment is $\mu\approx 10^{-19}\omega_{(5)}$~erg
gauss$^{-1}$ (where $\omega_{(5)}\equiv \omega/10^5{\rm s}^{-1}$). Therefore
the Larmor precession has a period 
$t_{Lar}\approx 3\times 10^6 B_{(5)}^{-1}$~s.

Nevertheless, suprathermal, thermal and subthermal grain rotation are just 
components of complex
grain dynamics. As any solid body, interstellar grains can rotate about 3 
different body axes. As the
result they tumble while rotating. This effect was attracting attention 
of the early researchers (see
Jones \& Spitzer 1967) till Purcell (1979) identified internal relaxation 
within grains as the process
that can suppress grain rotation about all axes, but the axis corresponding 
to the grain maximal
moment of inertial (henceforth axis of maximal inertia). Indeed,  consider 
a spheroidal grain, which
kinetic energy can be presented as (see Lazarian \& Roberge 1997) $
E(\theta)=\frac{J^2}{I_{max}}\left(1+\sin^2\theta (h-1)\right), $
where $\theta$ is the angle between the axis of major inertia and grain 
angular momentum.
In the absence of external torques grain angular momentum is preserved. 
The minimum of grain energy 
corresponds therefore to $\theta=0$, or grain rotating exactly about the 
axis of maximal inertia. As 
internal dissipation decreases kinetic energy, it sounds natural that
 $\theta=0$ is the expected 
state of grain subjected to fast internal dissipation.

Purcell (1979) introduced a new process of internal dissipation which he 
termed "Barnett relaxation". 
This process may be easily understood. We know that a 
freely rotating grain preserves the direction of
${\bf J}$, while angular velocity precesses about 
${\bf J}$ and in grain body axes.
We learned earlier that the Barnett effect results in the magnetization
vector parallel to $\vec \Omega$. As a result, the Barnett magnetization
will precess in body axes and cause paramagnetic relaxation.
The ``Barnett equivalent magnetic field'', i.e. the equivalent external
magnetic field that would cause the same magnetization of the grain  
material, is $H_{BE}=5.6 \times10^{-3} \omega_{(5)}$~G, 
which is much larger than the interstellar magnetic 
field. Therefore the Barnett relaxation happens on the scale $t_{Bar}\approx
4\times 10^7 \omega_{(5)}^{-2}$~sec,
i.e. essentially instantly compared to the time that it takes to
damp grain rotation for typical molecular cloud conditions.

Even stronger relaxation process has been identified recently by 
Lazarian \& Draine (1999a). They termed it ``nuclear relaxation''.
 This is an analog of Barnett
relaxation effect that deals with nuclei. Similarly to unpaired electrons
nuclei tend to get oriented in a rotating body. However the nuclear analog
of ``Barnett equivalent'' magnetic field is much larger and Lazarian \&
Draine (1999a) concluded that the nuclear relaxation can be a million times
faster than the Barnett relaxation. 

Why would the actual relaxation rate matter? The
rate of internal relaxation couples grain rotational and vibrational
degrees of freedom. LD99b showed that this
will result in grain ``thermal flipping''. Such a flipping would
average out Purcell's torques and result in grain being
``thermally trapped'' in spite of the presence of uncompensated
torques. Whether grain gets ``thermally trapped'' depends on
its size (with the grains less than a critical size $a_c$
rotating thermally). 
While Barnett and inelastic relaxation (see also Lazarian \& Efroimsky
1999) results in $a_c$ equal or less than $10^{-5}$~cm, the
nuclear internal relaxation provides $a_c\sim 10^{-4}$~cm. This means
that most grains rotate thermally in the presence of Purcell's torques.
The exception to this thermallization are radiative torques that are
not fixed in grain coordinates. Such torques can spin-up dust in spite
of thermal flipping.

\section{What does align grains?}

{\bf Paramagnetic Alignment}

Davis-Greenstein (1951)
mechanism (henceforth D-G mechanism)
is based on the paramagnetic dissipation that is experienced
by a rotating grain. Paramagnetic materials contain unpaired
electrons which get oriented by the interstellar magnetic field ${\bf B}$. 
The orientation of spins causes
grain magnetization and the latter 
varies as the vector of magnetization rotates
 in grain body coordinates. This causes paramagnetic loses 
at the expense of grain rotation energy.
Note, that if the grain rotational velocity ${\vec \Omega}$
is parallel to ${\bf B}$, the grain magnetization does not change with time
and therefore
no dissipation takes place. Thus the
paramagnetic dissipation  acts to decrease the component of ${\vec \Omega}$
perpendicular to ${\bf B}$ and one may expect that eventually
grains will tend to rotate with ${\vec \Omega}\| {\bf B}$
provided that the time of relaxation $t_{D-G}$ is much shorter than  $t_{gas}$,
the
time of randomization through chaotic gaseous bombardment.
In practice, the last condition is difficult to satisfy. For $10^{-5}$ cm 
grains
in the diffuse interstellar medium
$t_{D-G}$ is of the order of $7\times 10^{13}a_{(-5)}^2 B^{-2}_{(5)}$s , 
while  $t_{gas}$ is $3\times 10^{12}n_{(20)}T^{-1/2}_{(2)} a_{(-5)}$ s (
see table~2 in Lazarian \& Draine 1997) if
magnetic field is $5\times 10^{-6}$ G and
temperature and density of gas are $100$ K and $20$ cm$^{-3}$, respectively. 
However, at the time when it was introduced ,in view of uncertainties in
interstellar parameters, the D-G mechanism looked plausible.

The first detailed analytical treatment of the problem of D-G
alignment was given by Jones \& Spitzer (1967) who described the alignment
of ${\bf J}$
using a Fokker-Planck equation. This 
approach allowed them to account for magnetization fluctuations
within grain material and thus provided a more accurate picture of 
${\bf J}$ alignment. 
The first numerical treatment of
D-G alignment was presented by Purcell (1969). 
By that time it became clear that the D-G
mechanism is too weak to explain the observed grain alignment. However,
Jones \& Spitzer (1967) noticed that if interstellar grains
contain superparamagnetic, ferro- or ferrimagnetic (henceforth SFM) 
inclusions\footnote{The evidence for such inclusions was found much later
through the study of interstellar dust particles captured in
the atmosphere (Bradley 1994).}, the
$t_{D-G}$ may be reduced by orders of magnitude. Since $10\%$ of
atoms in interstellar dust are iron
the formation of magnetic clusters in grains was not far fetched
(see Martin 1995).
However, detailed calculations in Lazarian (1997), Roberge \& Lazarian
(1999) showed that the alignment achievable cannot account for
observed polarization coming from molecular clouds provided
that dust grains rotate thermally. This is the consequence of
thermal fluctuations within grain material. These internal
magnetic fluctuations
randomize grains orientation in respect to magnetic field if
grain body temperature is close to the rotational temperature.

Purcell (1979) pointed out that fast rotating grains are immune to
both gaseous and internal magnetic 
randomization. Thermal trapping limits the range of grain sizes
for which Purcell's torques can be efficient (Lazarian \& Draine 1999ab).
Grains with the sizes larger than the wavelength size can be spun up
by the incoming starlight, however (see Draine \& Weingartner 1996).

{\bf Mechanical Alignment}

Gold (1951) mechanism is a process of mechanical alignment of grains. Consider
a needle-like grain interacting with a stream of atoms. Assuming
that collisions are inelastic, it is easy to see that every
bombarding atom deposits angular momentum $\delta {\bf J}=
m_{atom} {\bf r}\times {\bf v}_{atom}$ with the grain, 
which is directed perpendicular to both the
needle axis ${\bf r}$ and the 
 velocity of atoms ${\bf v}_{atom}$. It is obvious
that the resulting
grain angular momenta will be in the plane perpendicular to the direction of
the stream. It is also easy to see that this type of alignment will
be efficient only if the flow is supersonic\footnote{Otherwise grains
will see atoms coming not from one direction, but from a wide cone of
directions (see Lazarian 1997a) and the efficiency of 
alignment will decrease.}.
Thus the main issue with the Gold mechanism is to provide supersonic
drift of gas and grains. Gold originally proposed collisions between
clouds as the means of enabling this drift, but later papers (Davis 1955) 
showed that the process could  only align grains over limited patches of
interstellar space, and thus the process
cannot account for the ubiquitous grain 
alignment in diffuse medium.

Suprathermal rotation introduced in Purcell (1979) persuaded researchers
that mechanical alignment is marginal. Indeed, fast rotation makes
it difficult for gaseous bombardment to align grains. However, two new
developments must be kept in mind. First of all, a number of papers
proved that mechanical alignment of suprathermally rotating grains
is possible (Lazarian 1995, Lazarian \& Efroimsky 1996, Efroimsky 2002). Moreover, recent work on
grain dynamics (Lazarian \& Yan 2002, Yan \& Lazarian 2003) proved
that MHD turbulence can render grains with supersonic velocities.
While we do not believe that mechanical alignment is the dominant
process, it should be kept in mind in analyzing observations
(see Rao et al. 1998).

{\bf Alignment via Radiative Torques}

Anisotropic starlight radiation can both spin the grains and align them.
This was first realized by Dolginov \& Mytrophanov (1976), but this
work came before its time. The researchers did not have reliable means
to study dynamics of grains and the impact of their work was marginal.
Before Bruce Draine realized that the torques
can be treated with the versatile discrete dipole approximation (DDA)
code ( Draine \& Flatau 1994) the radiative torque alignment was
very speculative. For instance, earlier on
difficulties associated with the analytical approach to
the problem were discussed in Lazarian (1995a).
However, very soon after that Draine (1996) modified the DDA code
 to calculate the torques acting on grains of arbitrary
shape. His work revolutionized the field! 
The magnitude of torques were found to be substantial and present
for grains of various irregular shape (Draine 1996, Draine \& Weingartner
1996). After that it became impossible
to ignore these torques.

One of the problem of the earlier treatment was that in the presence
of anisotropic radiation the torques will change as the grain aligns
and this may result in a spin-down.  Moreover,
anisotropic flux of radiation will deposit angular momentum 
which is likely to overwhelm rather weak paramagnetic torques. These sort of
questions were addressed by Draine \& Weingartner (1997) and it was
found that for most of the tried grain shapes the torques tend to 
align ${\bf J}$ along magnetic field. The reason for that is yet unclear
and some caution is needed as the existing treatment ignores the dynamics
of crossovers which is  very important for the alignment of
suprathermally rotating grains. A recent work by Weingartner \&
Draine (2003) treats flipping of grains in the presence of monochromatic
radiation. 

\subsection{What is Future Work?}

Observational testing of alignment is extremely important. 
Both the dependences of the polarization degree versus wavelength
that follow Serkowski law (Serkowski 1973) and 
studies
of changes of polarization degree with the wavelength done in Far Infrared
(see Hildebrand
2000) are consistent with theoretical predictions (see discussion in
Lazarian, Goodman \& Myers 1997). 
According to Lazarian (2003) the study of grain alignment
at the diffuse/dense cloud interface by Whittet et al. (2001)
is suggestive that grains there are being aligned by radiative torques. 

Radiative torques look as the most attractive mechanism to align grains
in molecular clouds. However more theoretical work is required to understand
why grains subjected to anisotropic radiation get preferentially
aligned with their long axes perpendicular to magnetic field. 

{\bf Acknowledgments.} I thank H. Yan for suggestions and help.
The research is supported by the NASA grant 0830 300 N665 736.

%
\centering
%
%

%


\printindex
\end{document}